\documentclass[10pt,twocolumn]{article}

\usepackage[margin=0.9in]{geometry}
\usepackage{graphicx}
\usepackage{amsmath}
\usepackage[utf8]{inputenc}
\usepackage{newunicodechar}
\usepackage{titling}
\usepackage{authblk}
\usepackage[]{abstract}
\usepackage[font=small,labelfont=bf]{caption}
\usepackage{titlesec}
\usepackage{float}

\titleformat*{\section}{\large\bfseries}
\titleformat*{\subsection}{\normalsize\bfseries}

\newunicodechar{¹}{$^{1}$}
\newunicodechar{³}{$^{3}$}
\newunicodechar{⁴}{$^{4}$}

\title{Dual-comb interferometry via repetition-rate switching of a single frequency comb}

\author[1,*]{David R. Carlson}
\author[1]{Daniel D. Hickstein}
\author[1,2]{Daniel C. Cole}
\author[1,2]{Scott A. Diddams}
\author[1,2]{Scott B. Papp}

\affil[1]{Time and Frequency Division, National Institute of Standards and
  Technology, 325 Broadway, Boulder, CO 80305, USA}
\affil[2]{Department of Physics, University of Colorado, 2000 Colorado Ave., Boulder, CO 80309, USA}

\affil[*]{Corresponding author: david.carlson@nist.gov}

\date{\vspace{-3mm}\small\today}

\makeatletter
\newbox\abstract@box
\renewenvironment{abstract}
  {\global\setbox\abstract@box=\vbox\bgroup
     \hsize=\textwidth\linewidth=\textwidth
    \small
    \vspace{-6mm}
    \quotation}
  {\endquotation\egroup}
\expandafter\def\expandafter\@maketitle\expandafter{\@maketitle
  \ifvoid\abstract@box\else\unvbox\abstract@box\if@twocolumn\vskip1.5em\fi\fi}
\makeatother

\makeatletter
\renewcommand{\fnum@figure}{Fig. \thefigure}
\makeatother

\begin{document}

\begin{abstract}
We experimentally demonstrate a versatile technique for performing dual-comb interferometry using a single frequency comb. By rapid switching of the repetition rate, the output pulse train can be delayed and heterodyned with itself to produce interferograms. The full speed and resolution of standard dual-comb interferometry is preserved while simultaneously offering a significant experimental simplification and cost savings. We show that this approach is particularly suited for absolute distance metrology due to an extension of the non-ambiguity range as a result of the continuous repetition-rate switching.
\end{abstract}

\maketitle

The interference between two mutually coherent optical frequency combs with slightly different repetition rates can be used to extract temporal and spectral information from the combs with both high resolution and high speed~\cite{schiller_spectrometry_2002,coddington_dual-comb_2016}.  This technique is known generally as dual-comb interferometry and has been successfully demonstrated for applications including spectroscopy~\cite{coddington_coherent_2008}, hyperspectral imaging~\cite{shibuya_scan-less_2017}, distance metrology~\cite{coddington_rapid_2009}, ellipsometry~\cite{minamikawa_dual-comb_2017}, and two-way time transfer~\cite{giorgetta_optical_2013}. The most common implementation uses two independent but optically phase-locked combs; though simplifications involving birefringent~\cite{link_dual-comb_2017}, bidirectional~\cite{mehravar_real-time_2016,ideguchi_kerr-lens_2016}, or spatially multiplexed~\cite{lucas_spatial_2018} resonators are possible, albeit with limited tunability in the repetition rates. Alternatively, a single comb source can be used to obtain dual-comb interferograms through continuous repetition-rate sweeping~\cite{hochrein_optical_2010}, or by using an acousto-optic programmable dispersive filter to continuously shift the delay of a pulse train~\cite{znakovskaya_dual_2014}.  However, to achieve these simplifications, these techniques must sacrifice either acquisition speed or resolution compared to standard methods.

Here we demonstrate a simple and versatile new approach to dual-comb interferometry, using a single comb laser, that relies on rapid switching of the pulse repetition rate to produce interferograms at the full speed and resolution of standard dual-comb methods. We call this technique \textbf{P}arallel \textbf{H}eterodyne \textbf{I}nterferometry via \textbf{R}ep-rate \textbf{E}xchange (PHIRE) and show how it can be used for applications including spectroscopy, vibrometry, and absolute distance metrology. 

\begin{figure}
\includegraphics[width=\linewidth]{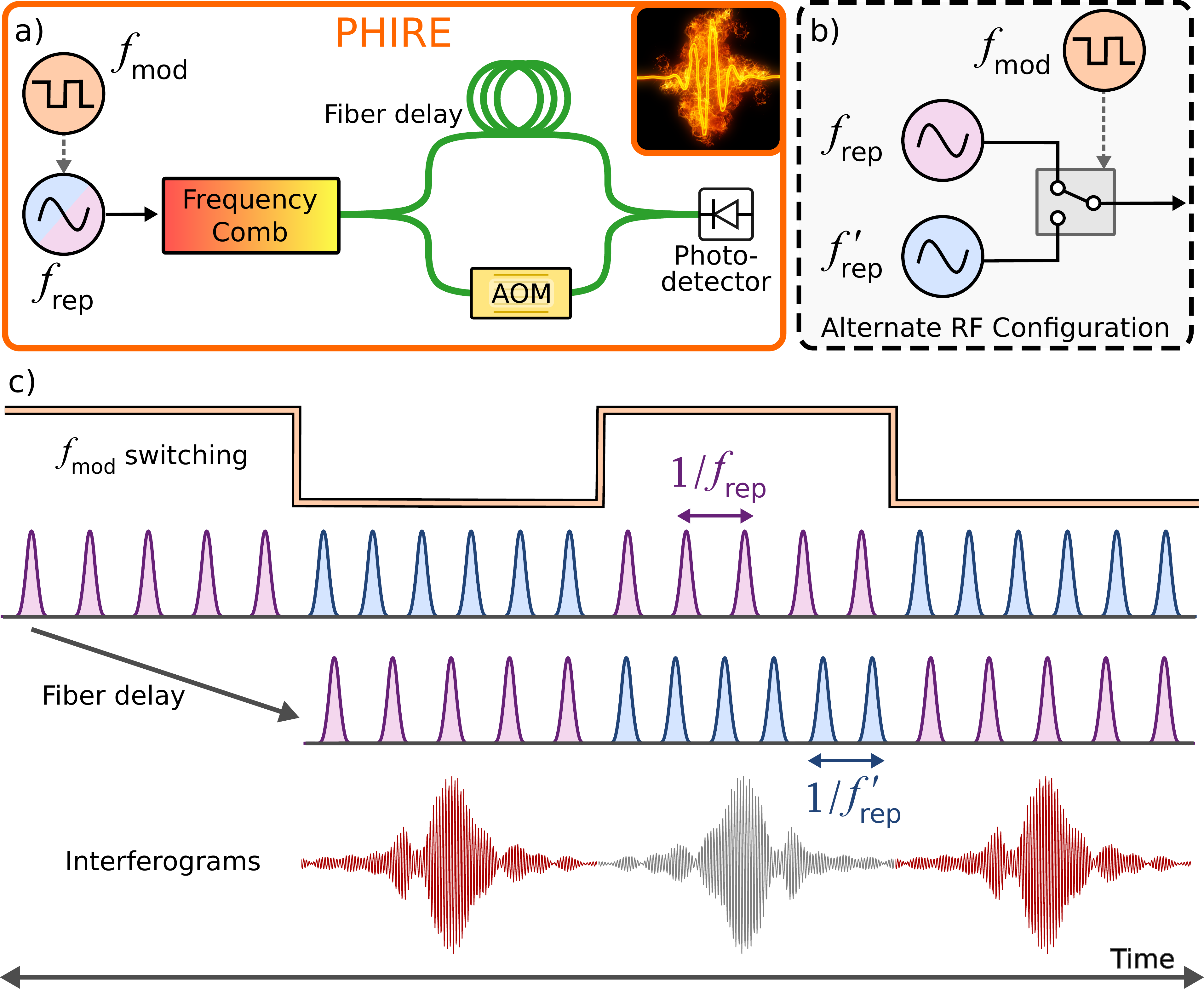}
\caption{Concept of Parallel Heterodyne Interferometry via Rep-rate Exchange (PHIRE). a) An RF source providing the comb repetition rate $f_\textrm{rep}$ is periodically modulated to produce two distinct frequencies that alternate at a rate of $f_\textrm{mod}$. The comb output is split and recombined with itself after passing through a fixed length of optical fiber to produce dual-comb interferograms. An acousto-optic modulator (AOM) provides a frequency offset for one arm to shift the interferogram carrier frequency away from baseband. b) An alternate RF configuration using electronic switching to toggle between two different sources for $f_\textrm{rep}$. c) The modulation $f_\textrm{mod}$ is matched to the length of the fiber delay such that the photodetected output produces interferograms from the two overlapping pulse trains with repetition rates $f_\textrm{rep}$ and $f_\textrm{rep}^\prime$.}
\label{fig:concept}
\end{figure}

The fundamental PHIRE concept is illustrated by the schematic in Fig.~\ref{fig:concept}a.  A binary modulation $f_{\rm{mod}}$ is applied to an ultrafast laser to produce a pulse train whose repetition rate periodically alternates between $f_\textrm{rep}$ and $f_\textrm{rep}^\prime$, resulting in the creation of a time-multiplexed frequency comb. A fiber splitter then diverts half of the comb power to a delay line of length $L_{\rm{delay}}$ that is matched to half of the switching rate via $L_{\rm{delay}} = v_{\rm{g}}/(2f_{\rm{mod}})$, where $v_{\rm{g}}$ is the group velocity of the fiber at the comb's center wavelength. The other half of the comb power passes through an acousto-optic modulator (AOM) to provide a frequency offset for the photodetected RF spectrum, avoiding spectral aliasing at baseband frequencies. When these two arms are recombined, a sequence of interferograms is produced at the difference in repetition rates $\Delta f_{\rm{rep}}$ of the two pulse trains. If absolute calibration of the comb spacing or support for coherent time-domain averaging is desired, $f_\textrm{mod}$ can alternatively be used to select between two phase-coherent RF sources via an RF switch, as shown in Fig.~\ref{fig:concept}b.  However, in both configurations, the interferograms are reversed every half-cycle of $f_{\rm{mod}}$ due to the periodic swapping of pulse trains in each arm of the interferometer. Unlike standard dual-comb interferometry, this means that acquisitions longer than $1/(2f_\textrm{mod})$ require extra processing in order to avoid spectral artifacts due to the out-of-phase interferograms. Nevertheless, interferogram sequences shorter than $1/(2f_\textrm{mod})$ can be digitized and Fourier transformed directly to yield the full phase and amplitude spectrum of the comb.

The length of the fiber delay determines the optimum value of $f_\textrm{mod}$ and thus also limits what values of $\Delta f_{\rm{rep}}$ are supported with full frequency resolution. For example, when $\Delta f_\textrm{rep} < 2f_\textrm{mod}$, interferograms are acquired at the higher rate of $2f_\textrm{mod}$, though resolution is reduced due to the truncated delay range. This scenario can be useful when acquisition speed takes priority over resolution, but it may be necessary to control the relative phase of the RF sources to center the interferogram within the scan window.  On the other hand, when $\Delta f_\textrm{rep} > 2f_\textrm{mod}$, multiple interferograms will be acquired per half-period, with a maximum value limited only by aliasing at the Nyquist-zone edges for a given spectral bandwidth. This sequence of multiple interferograms can be Fourier transformed to yield a comb-resolved spectrum, or processed individually and averaged to improve the signal-to-noise ratio of a single interferogram.

While the PHIRE technique can be applied to any type of ultrafast pulse source capable of quickly changing repetition rate, we demonstrate its implementation using an electro-optic-modulation (EOM) comb.  EOM combs are simple and robust sources of pulses that are carved directly from a CW laser through cascaded phase and intensity modulation~\cite{kobayashi_highrepetitionrate_1972,kourogi_wide-span_1993,carlson_ultrafast_2017}. Because this pulse generation is driven directly by an electronic oscillator, the resulting combs have an inherently high degree of tunability and flexibility in their configuration. Moreover, they are particularly suited for dual-comb spectroscopy because each comb can be derived from the same CW laser, ensuring a high degree of mutual coherence~\cite{duran_ultrafast_2015,martin-mateos_dual_2015,millot_frequency-agile_2015,fleisher_coherent_2016}.

\begin{figure}
\includegraphics[width=\linewidth]{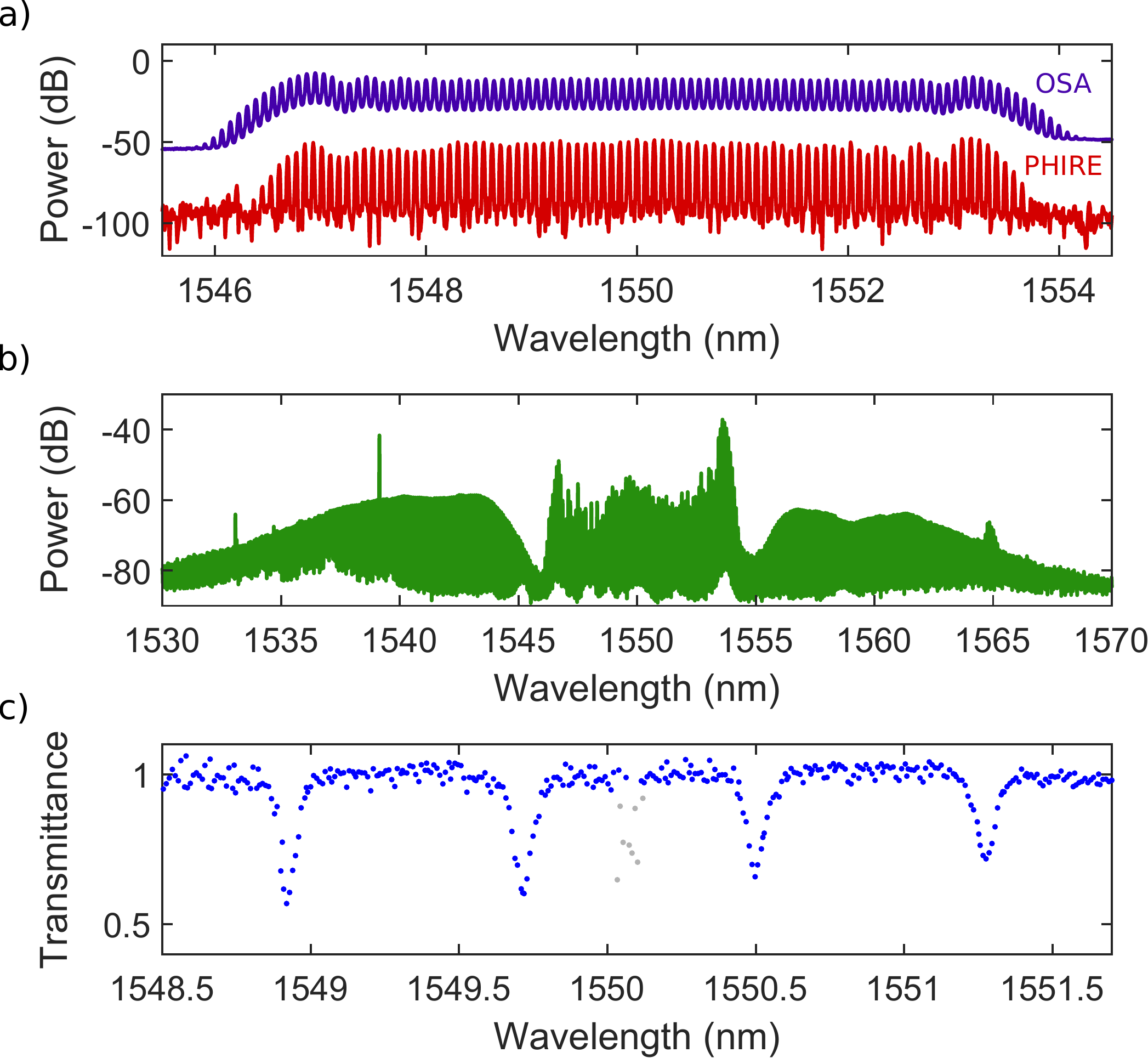}
\caption{a) Electro-optic comb spectrum obtained with an optical spectrum analyzer (OSA, purple) and via PHIRE (red). b) PHIRE spectrum retrieved after broadening the initial time-multiplexed pulse trains in highly nonlinear fiber. c) Absorption spectroscopy of H$^{14}$C$^{13}$N gas around the comb's center wavelength. Comb spectrum scanned in steps of approximately 0.01 nm. Gray points are the result of laboratory technical noise at the AOM operating frequency.}
\label{fig:spectrum}
\end{figure}

To demonstrate operation of the PHIRE system, we use an EOM comb with a nominal repetition rate of 10.1~GHz that implements the RF configuration from Fig.~\ref{fig:concept}b. A difference in repetition rates of $\Delta f_\textrm{rep} = 3.28$~MHz is set by the two RF synthesizers and is chosen to be an integer multiple of $f_\textrm{mod} = 164$~kHz, which is matched to the $\sim$600~m fiber delay. This configuration allows up to ten interferograms to be acquired during each switching cycle and supports comb-tooth-resolved spectra. In our implementation, the optical spectrum is mapped to the RF domain with the comb's center wavelength corresponding to the AOM drive frequency of 250~MHz. For comparison, Fig.~\ref{fig:spectrum}a shows the comb spectrum obtained via PHIRE from a 16$\times$ average of 2.8-$\mu$s time-domain traces versus that obtained with a high-resolution optical spectrum analyzer (OSA). In addition to having higher resolution, the acquisition speeds with PHIRE can be up to four orders of magnitude faster than the OSA.

Increasing the comb bandwidth is possible by undergoing nonlinear spectral broadening before passing through the delay line.  In fact, the time-multiplexing inherent to PHIRE is ideal for this type of process because the pulses do not walk across each in the nonlinear medium and thus do not introduce higher-order nonlinearities in the dual-comb interferograms. An example nonlinear PHIRE spectrum is shown in Fig.~\ref{fig:spectrum}b, obtained after the comb has been broadened in 4~m of highly nonlinear fiber. To maintain comb-tooth-resolution, $\Delta f_\textrm{rep}$ was reduced to 1~MHz while $L_\textrm{delay}$ was increased to 3~km. However, a practical limitation for nonlinear PHIRE using an electro-optic source is that the two resulting combs have slightly different spectral bandwidths, and thus a slightly different pulse chirp is required to optimally compress them~\cite{kobayashi_optical_1988}. However, it should be possible to use electronically controlled RF phase shifters to adjust the output pulse chirp synchronously with $f_\textrm{mod}$ to mitigate this effect and allow stronger nonlinear interactions, possibly supporting direct self-referencing~\cite{beha_electronic_2017,carlson_ultrafast_2017}.

As in other forms of frequency comb spectroscopy, molecular absorption measurements can be performed across the comb bandwidth. For example, Fig.~\ref{fig:spectrum}c shows several absorption bands in H$^{14}$C$^{13}$N gas, obtained after passing through a 15-cm pure gas cell at a pressure of 100~Torr and subtracting the background~\cite{gilbert_hydrogen_1998}. To increase the effective resolution of the PHIRE measurement, we leverage the tunability of the CW pump laser to sweep the entire comb across the absorption features in steps of $\sim$0.01~nm. Absolute calibration of the wavelength in this case would require self-referenced stabilization; however many applications for high-repetition-rate combs don't require such accuracy.

While fast acquisition speeds and a simple experimental setup are important features of PHIRE, there are also tradeoffs to consider in a real implementation.  First, compared to conventional dual-comb interferometry with EOM sources, the noise requirements for the common CW pump laser are more demanding because the coherence time of the laser must exceed the propagation time through the fiber.  However, this is easily achieved with many single-frequency fiber and diode lasers having linewidths less than 100~kHz. Second, the fiber delay can introduce slow phase fluctuations that may limit the ability to perform long-duration coherent averaging. However, for applications sensitive to this drift, it is possible to achieve a high-degree of path-length stabilization by moving the AOM to the delay arm of the interferometer and then adding a fiber to ``shortcut'' the delay in order to generate a heterodyne beat at the AOM frequency that can be actively locked. Finally, for broad spectral bandwidths, higher-order dispersion in the long fiber delay may lead to poor temporal overlap across the full spectral range unless special care is taken to compensate for this effect.

\section*{Application: ranging and vibrometry}
When applied to absolute distance metrology, PHIRE has a unique advantage compared to other dual-comb systems by easily extending the maximum measurable distance, known as the non-ambiguity range (NAR). In a typical dual-comb ranging experiment, the NAR for a time-of-flight (TOF) analysis is given as $\textrm{NAR} = v_{g}/(2f_\textrm{rep})$. Consequently, for high-repetition-rate combs, the fast acquisition speeds capable of resolving rapidly moving objects come at the price of a reduced NAR~\cite{trocha_ultrafast_2018}.

Fortunately, previous dual-comb ranging work has shown that by changing the repetition rate of the comb, the measurement NAR can be increased by several orders of magnitude via the Vernier effect~\cite{coddington_rapid_2009,lee_absolute_2013}. However, with PHIRE this extension of the NAR happens automatically due to the alternating values of  $f_\textrm{rep}$, provided the interferograms from each half-cycle of $f_\textrm{mod}$ are processed separately.  The result is that with the 10.1-GHz comb used above, the NAR can be extended from 1.5~cm to  $\textrm{NAR}_\textrm{ext} = v_{g}/(2\Delta f_{\rm{rep}}) = 45.7$~m under atmospheric conditions when $\Delta f_\textrm{rep} = 3.28$~MHz. 

\begin{figure}
\includegraphics[width=\linewidth]{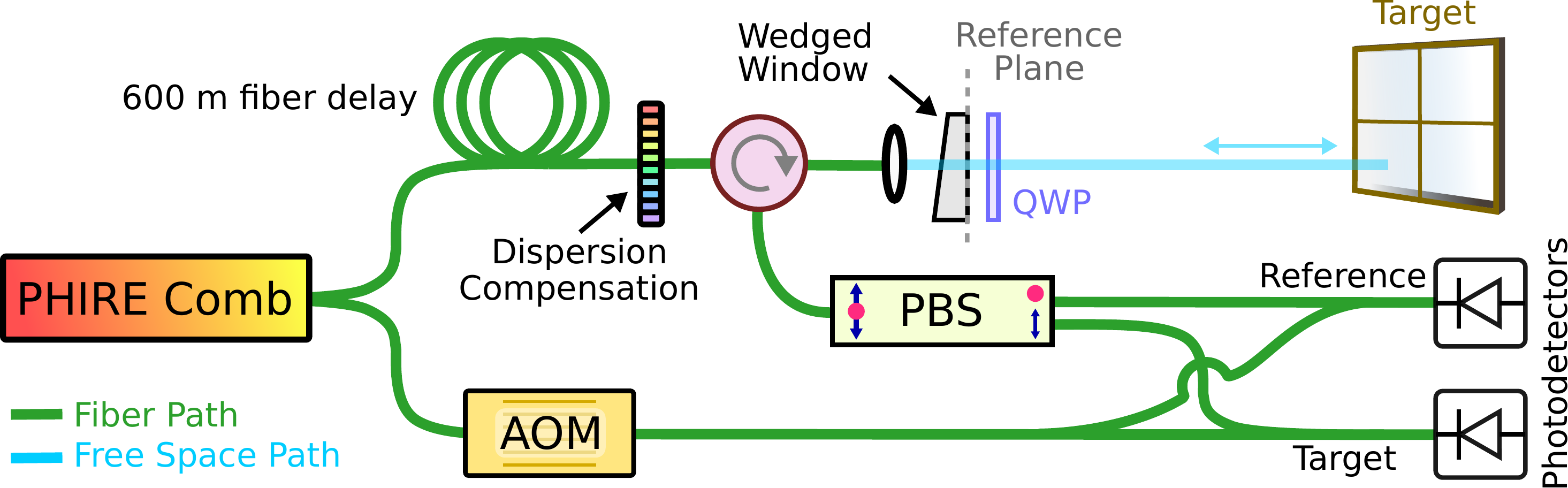}
\caption{Schematic for ranging and vibrometry measurements with PHIRE. A wedged glass window is used to isolate a single surface reflection to serve as the reference plane from which optical path lengths are measured. A quarter-wave plate (QWP) between the reference surface and the target allows orthogonal polarizations to be used for the two interferogram sources, avoiding measurement dead zones. AOM, acousto-optic modulator; PBS, polarization beam splitter.}
\label{fig:ranging}
\end{figure} 

In our experimental configuration (Fig.~\ref{fig:ranging}), we use separate detectors for the target and reference interferograms in order to avoid the measurement dead zones that occur with a single detector~\cite{coddington_rapid_2009}.  Splitting of the two interferogram signals is achieved via orthogonal polarization states by inserting a quarter-wave plate between the reference window and the target~\cite{lee_absolute_2013}. As a result, an additional calibration step must be taken to compensate for the different optical path lengths traversed by the two beams upon separation by a polarizer. If necessary, the precise thickness of the wave plate can also be characterized separately and included in the final calculation for absolute measurements.

Interferograms from the two channels are then collected with an oscilloscope at a sampling rate of 2~GSa/s and then transferred to a computer for further processing. After resampling to a rate of $610\times\Delta f_\textrm{rep}=2.0008$~GSa/s, the acquisition is divided into shorter segments corresponding to each half cycle of $f_{\rm{mod}}$ and truncated to contain exactly eight interferograms.  Each segment is then Fourier transformed, and the spectral phase of each tooth is obtained and unwrapped across the comb bandwidth. The unwrapped phases for both the target and reference channels are then subtracted and a linear fit is performed on the difference. The resulting phase, as a function of optical frequency $\nu$, is of the form $\phi = \phi_0 + [\textrm{d}\phi/\textrm{d}\nu]\nu$, where the absolute phase $\phi_0$ contains the interferometric data and the slope $\textrm{d}\phi/\textrm{d}\nu$ encodes the TOF distance through $L_\textrm{TOF} = (c/4\pi)[\textrm{d}\phi/\textrm{d}\nu]$, with $c$ being the speed of light.

\begin{figure}
\includegraphics[width=\linewidth]{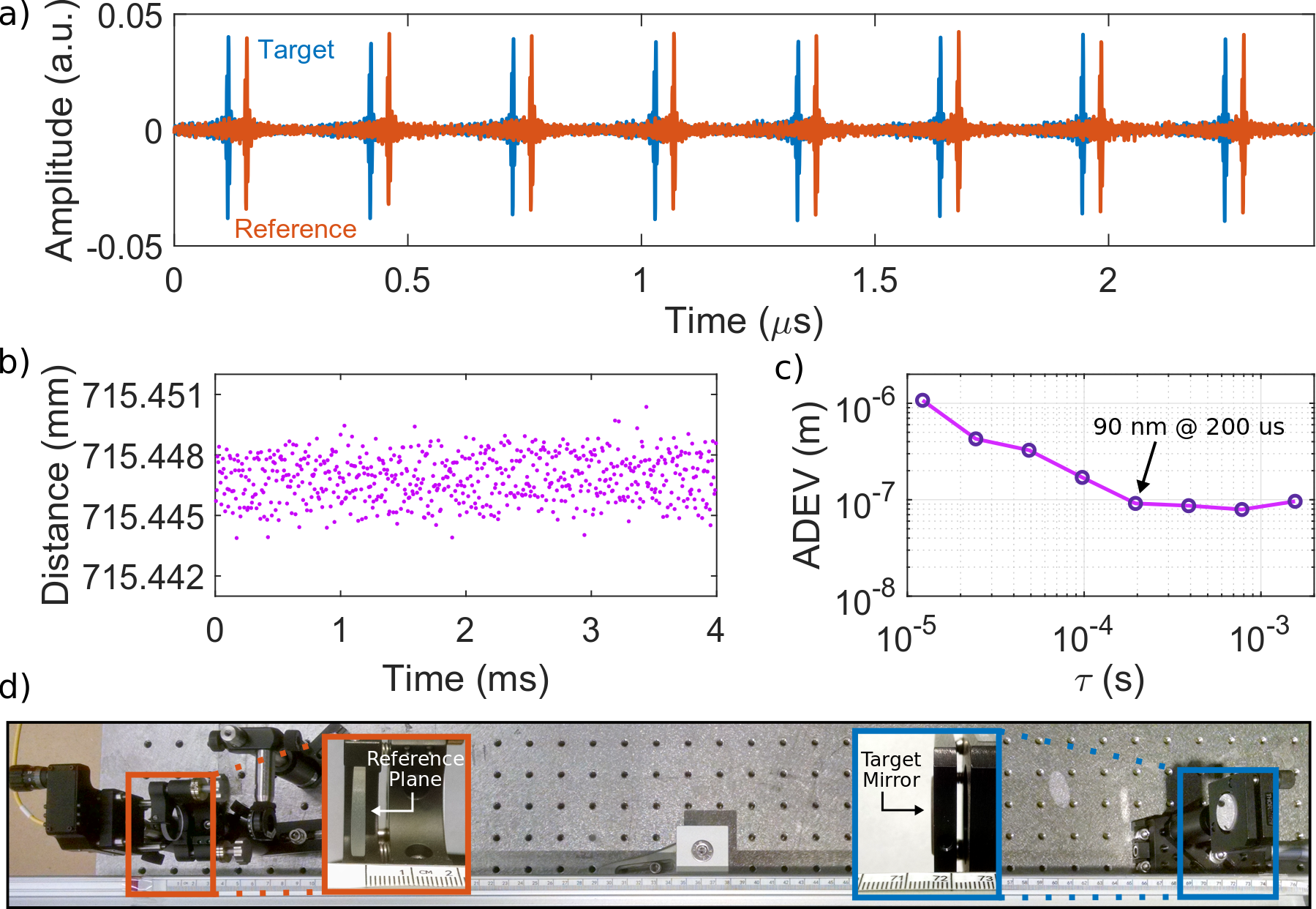}
\caption{a) Interferogram sequence for a single half-period of $f_{\rm{mod}}$ from both the reference and target detectors. b) Extracted time-of-flight absolute distance data. Each point represents a single half-period of $f_{\rm{mod}}$, containing eight individual interferograms. c) Allan Deviation (ADEV) of the extracted time-of-flight distance data. d) Photograph of the experiment confirming the correct non-ambiguity-range multiple was determined.}
\label{fig:absdistance}
\end{figure} 

To determine the precise NAR multiple $m$ by which the absolute distance exceeds the measured value, interferograms from both half-cycles of $f_\textrm{mod}$ are processed simultaneously to provide two slightly different distances, $L_\textrm{TOF}$ and $L_\textrm{TOF}^\prime$. With the corresponding non-ambiguity ranges, the multiple $m$ can then determined by $ m = (L_\textrm{TOF} - L_\textrm{TOF}^\prime) / (\textrm{NAR} - \textrm{NAR}^\prime)$.  However, because separate detectors are used for the reference and target channels, an additional measurement must be taken to calibrate the differential path length. To do this, the beam traveling to the target is blocked and a fiber polarization controller adjusts the polarization of the reference pulse to equally split between polarization states. In our case, we were able to determine the NAR multiple of the calibration to be $m=103$ and the air-equivalent distance to be $L_\textrm{cal} = 1.530$~m. When the target beam is unblocked, the same procedure can be used to retrieve the absolute distance after subtracting the calibration measurement via $L_\textrm{abs} = L_\textrm{cal} - (m\times\textrm{NAR} + L_\textrm{TOF})$. The result is shown in Fig.~\ref{fig:absdistance}b, with the correct NAR multiple ($m = 55$) verified by the photograph in Fig.~\ref{fig:absdistance}d. If careful consideration is given to the atmospheric parameters, an accuracy of one part in 10$^7$ should be achievable for the absolute distance~\cite{bobroff_recent_1993}.

The TOF measurement yields a precision of 1~$\mu$m after a single half-period of $f_\textrm{mod}$ and 90~nm after 200~$\mu$s of averaging, as shown in Fig.~\ref{fig:absdistance}c. We note that, despite the long fiber delay required for PHIRE, path-length fluctuations in the delay arm are entirely common mode for the reference and target pulses.  As a result, measurement precision is only limited by thermal drifts and acoustic noise in the free-space ranging segment and the unoptimized differential paths in the detection setup. 

Finally, to demonstrate the interferometric sensitivity of PHIRE, we perform a vibrometry measurement using the absolute phase $\phi_0$ of the linear fit to extract acoustic information from the sub-wavelength motion of a distant vibrating surface. With an EOM comb, the minimum sensitivity of this type of measurement is limited on short time scales by the phase stability of the CW pump laser, but can readily achieve sub-nm levels~\cite{teleanu_electro-optic_2017}.  

\begin{figure}
\includegraphics[width=\linewidth]{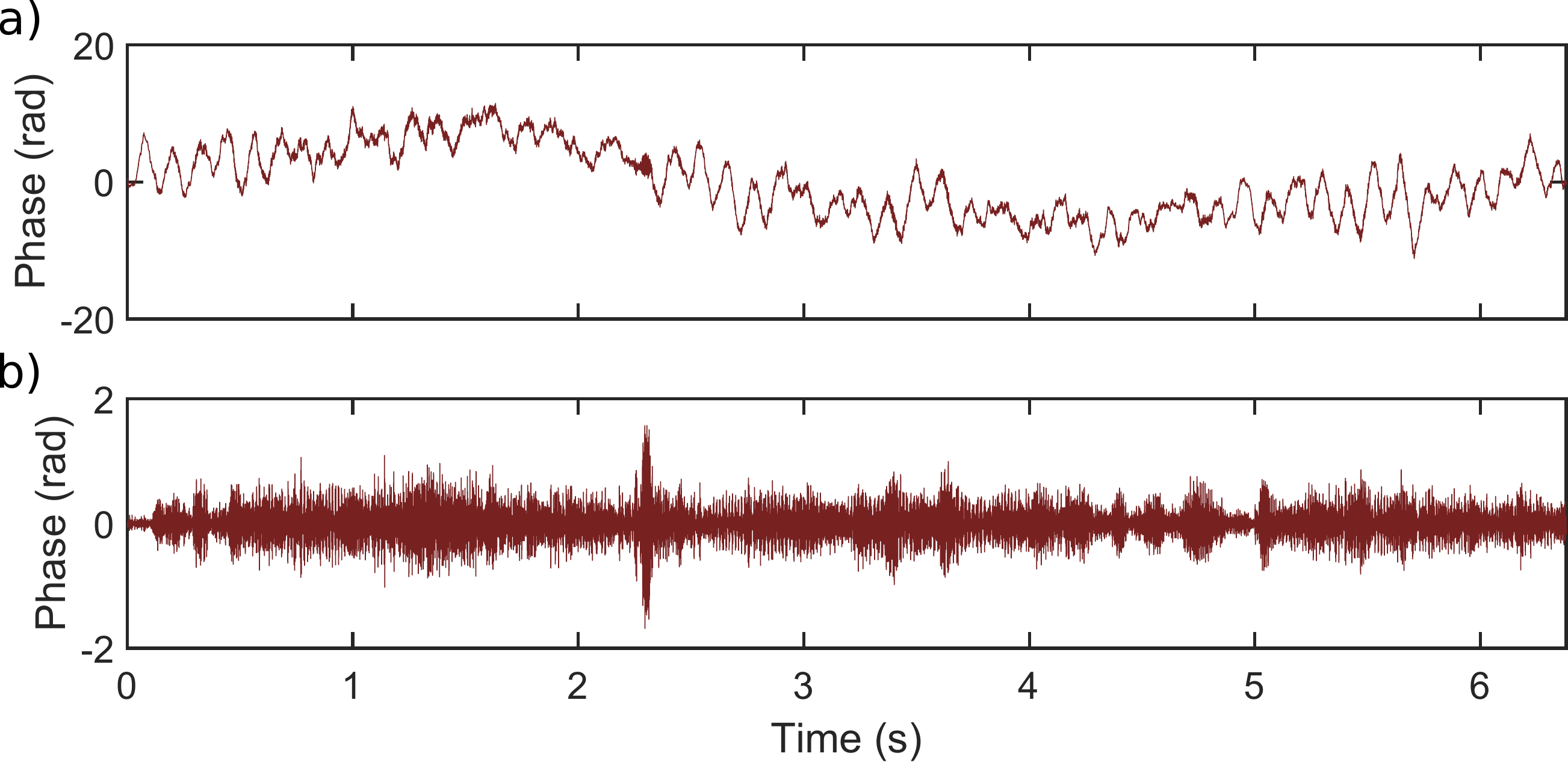}
\caption{At-a-distance acoustic eavesdropping in a noisy environment. a) Retrieval of the interferometric phase from the dual-comb acquisition is dominated by low-frequency optical path fluctuations. b) Applying a 150-Hz high-pass filter allows recovery of an intelligible waveform. Six seconds of extracted audio can be heard in Supplementary Media File 1.}
\label{fig:audio}
\end{figure}

In this example, the Fresnel reflection from a 1-m$^2$ plexiglass window approximately 4.5~m away serves as an eavesdropping target. An audio sample~\cite{cash_ring_1963} played through a small speaker at a typical listening volume 1~m behind the window can then be reconstructed directly from the interferometric phase data. Interferograms are saved in a segmented acquisition at a rate of 10~kHz, with each segment consisting of eight consecutive interferograms and yielding a single phase value that, when combined with all other segments, constructs the sampled audio waveform. Fig.~\ref{fig:audio}a shows the absolute interferogram phase as a function of time for this measurement. The strongest frequency components are at very low frequencies, and arise largely from slow path length fluctuations, either in the atmosphere or in the differential fiber path lengths.  Nevertheless, applying a single 150-Hz high-pass filter (Fig.~\ref{fig:audio}b) is sufficient to retrieve an intelligible audio waveform from the noisy background, as revealed in Supplementary Media File~1. 

Due to its simplicity, cost savings, and flexibility, PHIRE may allow dual-comb techniques to reach a wider variety of users and applications. Moreover, it should be possible to apply the technique to other types of pulsed lasers, and could be especially suited for combs whose repetition rates are controlled electronically, such as some semiconductor lasers~\cite{quinlan_harmonically_2009} and microresonator combs~\cite{obrzud_temporal_2017,cole_kerr-microresonator_2018}.

\vspace{6pt}
\noindent\large\textbf{Acknowledgment.} \normalsize The authors thank E. Baumann and F. Giorgetta for comments on the manuscript and I. Coddington for helpful discussions. This research is supported by the AFOSR under award number FA9550-16-1-0016, DARPA, NASA, NIST, and the NRC. This is a contribution of the U.S. government and is not subject to copyright.

\vspace{-6pt}
\small
\bibliography{Zotero}


\end{document}